\documentclass[aps,prl,twocolumn,showpacs]{revtex4-1}

\usepackage{amsmath}
\usepackage{amssymb}

\usepackage{graphicx}

\begin{document}

\title{Phase Description of Stochastic Oscillations}

\author{Justus T.~C.~Schwabedal}
\email{jschwabedal@googlemail.com}
\author{Arkady Pikovsky}
\affiliation{Department of Physics and Astronomy, Potsdam University, 14476 Potsdam, Germany}

\begin{abstract}
We introduce an invariant phase description of stochastic oscillations by
generalizing the concept of standard isophases.  The average
isophases are constructed as sections in the state space,
 having a constant mean first return time. The approach allows to obtain a global phase
variable of noisy oscillations, even in the cases where the
phase is ill-defined in the deterministic limit. A simple numerical method for finding the isophases is 
illustrated for noise-induced switching between two coexisting
limit cycles, and for noise-induced oscillation in an excitable system.  We also discuss how to determine 
the isophases for experimentally observed irregular oscillations, providing a basis for a refined phase description of observed oscillatory dynamics. 
\end{abstract}

\maketitle

Phase reduction is the basic tool in the characterization of self-sustained, autonomous oscillators.  
With a reasonably defined
phase variable, one obtains a one-dimensional representation of the
oscillator, allowing to describe important aspects of
its dynamics, such as regularity, sensitivity
to forcing and noise, etc~\cite{Winfree1980,Kuramoto1984,Pikovsky2001}. 
Furthermore, the concept of phase reduction is
substantial for the data analysis of oscillatory processes in
physics, chemistry, biology, and technical applications, where
various approaches exist for extracting 
phases from oscillatory time series~\cite{Schaefer-Rosenblum-Kurths-Abel-98,Rosenblum2001,Tokuda07,Bartsch2007,Stankovski12}.

To understand many properties of oscillating systems, such as their phase resetting and synchronizability, it is important to define the phases not
only for
the purely periodic motion, but for the whole state space. In the theory 
of deterministic oscillations this is done via so-called isochrones~\cite{Guckenheimer1975}, by attributing to a state with an arbitrary amplitude the phase of a point on the limit cycle, 
to which this state asymptotically converges. In this letter we generalize this concept to irregular, noisy oscillators. The main idea is based on 
the definition of the \textit{isophases} by virtue of the mean first
 passage time concept. We will first apply our method to noise-perturbed 
 deterministic oscillators for which 
the isophases can be compared with deterministic isochrones. Furthermore, 
we will
consider examples for which the isochrones and 
even the oscillations themselves disappear in the deterministic limit.
The method will be also applied 
to noise-perturbed chaotic oscillations. Finally, we will demonstrate its 
applicability to experimentally observed time series.

We start by reminding the standard definition of isophases (which in this 
case are also isochrones) in deterministic
systems with a stable limit cycle $\mathbf{x}_0(t)=\mathbf{x}_0(t+T)$ 
having period $T$. First, one defines the phase on the limit cycle 
$\varphi(\mathbf{x}_0)$. Then, being observed stroboscopically with time interval $T$, 
all the points $\mathbf{x}$ that converge to
a particular point on the cycle
$\mathbf{x}_0$  having
the  phase $\varphi(\mathbf{x}_0)$. These points form a Poincar\'e surface 
of section $J(\varphi(\mathbf{x}_0))$ for
the trajectories of the dynamical system, with the special property that 
the return time to this surface equals $T$ for all points on it. Thus, to find an 
isophase surface is equivalent to find a Poincar\'e surface of section with 
the
constant return time $T$.

For a noisy system we define the isophase surface $J$ as a
Poincar\'e surface of section, 
for which the mean first return time $J\to J$,
 after performing one full oscillation, is a constant $T$, having a 
meaning
of the average oscillation period.  In order for isophases to be 
well-defined, 
oscillations have to be well-defined as well: amplitudes should always 
remain positive (so that one can reliably recognize a ``full oscillation'').  
Otherwise the concept of phase 
is not well-defined for a random process, and the isophases are not 
meaningful. 

Analytical calculations of the mean first return time (MFRT) is a 
complex problem in dimensions larger than one, therefore below we apply a 
simple numerical algorithm for construction of the isophases:   
an initial Poincar\'e section
is iteratively altered until all mean return times are approximately equal.
In two-dimensional systems for which isophases are lines,
we represent Poincar\'e sections by a linear interpolation in
between a set of knots.  For each knot $x_j$, the average return 
time $T_j$ is computed via the Monte Carlo simulation.  According to the 
mismatch of
$T_j$ and the mean period $\langle T\rangle$, the knot $x_j$ is advanced or 
retarded.  The
procedure is repeated with all knots, until it converges and
all return times $T_j$ are nearly equal to
$\langle T\rangle$.

Before proceeding to different applications of this procedure, we discuss 
the importance of knowing isophases for noisy oscillations. The first 
important application is that of phase resetting.  Phase resetting curve 
determines how an oscillator responds to an external kick, this response 
determines synchronization properties of 
oscillator~\cite{Ermentrout2006,Achuthan-Canavier-09}. For deterministic 
oscillators the phase response curve is determined just from the isochrone 
to which the kick shifts the state of the system from the limit cycle. For 
irregular oscillators the proper definition of the phase response curve 
is based on the first passage time~\cite{Schwabedal2010b}, so to determine 
it one has to find to which isophase, as defined above, the system is 
shifted by the kick (note that in the limit of small noise, perturbative 
approaches for the phase dynamics do the 
job~\cite{Teramae2009,Goldobin2010}). The second application is in the 
analysis of experimental data of coupled oscillators 
(cf.~\cite{Schaefer-Rosenblum-Kurths-Abel-98,%
Rosenblum2001,Tokuda07,Bartsch2007,Stankovski12}). 
There one needs to determine the phase dynamics from the time series, this 
task is relatively simple to accomplish if the variations of the amplitudes 
are very small so that the definition of a phase-like variable 
along the observed limit cycle is unambiguous. However, in the presence of 
large irregular amplitude variations, the phase characterization of the 
oscillations is not unique (cf.~Fig.~\ref{fig:resp} below). 
Proceeding according to the given above 
definition of the isophases as the lines on the two-dimensional embedding 
plane, for which the mean return times do not depend on
the amplitudes, allows us to 
get rid of the ambiguity and to determine the phase in a consistent way.

\begin{figure}
\centering		
\includegraphics[width=0.8\columnwidth]{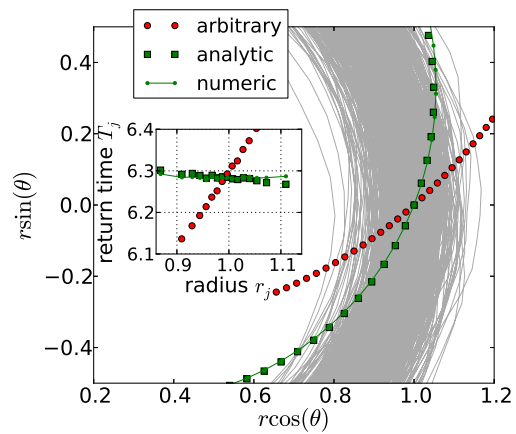}
\caption{ (color online)  The average isophase for noise-driven oscillator 
[Eq.~\ref{eq:ornUhlLS}] with $\omega=1,\kappa=1,\sigma=0.15,\gamma=1$ in 
cartesian coordinates.   The inset shows the reduction of the variations of 
the MFRT compared to an arbitrary cross-section.  The numerically-derived 
isophase shows minor differences to the analytic approximation 
[Eq.~\ref{eq:corrPhaseCorrection}]. }
	\label{fig:colored-plt-poincare}
\end{figure}

We stress here that in our definition of the isophases we do not assume 
Markovian property of the process, if the dynamics is non-Markovian, then 
the definition of the MFRT includes averaging over the ``prehistory'' or 
hidden variables as well. To illustrate this we consider as the first 
example a simple Stuart-Landau oscillator (variables $r,\theta$) perturbed 
by an Ornstein-Uhlenbeck noise (OUN) $\zeta(t)$:
\begin{equation}\begin{aligned}
	\dot{r} &= r(1-r^2)+\sigma r\zeta(t)\;,\qquad \dot\theta=\omega-\kappa (r^2-1),\\
	\gamma\dot{\zeta}&=-\zeta+\sqrt{\gamma}\xi(t)~,
\end{aligned}\label{eq:ornUhlLS}\end{equation}
where $\xi(t)$ denotes a $\delta$-correlated white noise, $\gamma$ is the 
correlation time of the OUN,  $\omega$ is the frequency of the noise-free 
limit cycle, and $\kappa$ is a nonisochronicity parameter. In the state 
space $r,\theta,\zeta$ the process is Markovian, but on the two-dimensional 
plane $r,\theta$ it is not. Nevertheless, by the method described we obtain 
numerical isophase for which the MFRT is nearly constant 
(Fig.~\ref{fig:colored-plt-poincare}). This isophase can be obtained also 
from the following analytic approximation. First, we introduce a 
``corrected'' phase variable $\psi=\theta-\kappa \ln r$ which obeys 
$\dot\psi=\omega+\sigma\kappa(\gamma\dot\zeta-\sqrt{\gamma}\xi(t))$. 
Averaging this expression and identifying $\omega=\dot\varphi$ where 
$\varphi$ is the correct uniformly rotating phase, we obtain 
$\varphi=\psi-\sigma\kappa\gamma\zeta$. In this expression we have to 
account for correlations of $\zeta$ and $r$, to obtain the isophases on the 
plane $(\theta,r)$. Assuming that $r$ follows $\zeta(t)$ adiabatically, we 
obtain $\sigma\zeta\approx r^2-1$, what leads to the following expression 
for the isophases
 \begin{equation}
	\varphi=\theta-\kappa\ln r-\kappa\gamma\left(r^2-1\right)~.
	\label{eq:corrPhaseCorrection}
\end{equation}
Isophase following from this formula is compared with numerical one in 
Fig.~\ref{fig:colored-plt-poincare}. Interestingly, the noise-induced 
correction (last term in (\ref{eq:corrPhaseCorrection})) does not contain 
parameter $\sigma$, but the range where this correction is valid $|r-1|\lesssim \sigma$ shrinks with the noise amplitude.

\begin{figure}[h]
\centering
\includegraphics[width=0.8\columnwidth]{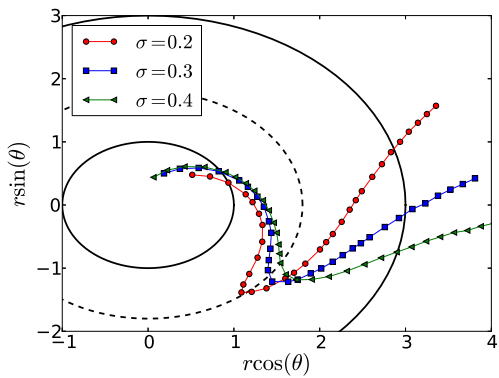}
	\caption{ (color online) Isophases of bistable oscillations mixed 
	by noise, in model (\ref{eq:bstc}) with $\omega=3, c=1.8,\delta=0$ and different noise intensities. Solid red curves are stable cycles, the dashed curve is the unstable cycle representing the basin boundary.}
	\label{fig:doubleLS-cmp-isoph}
\end{figure}

While in the simplest example above the effect of noise is in the 
correction of the deterministic isochrones only, we consider now a 
situation where local isophases of different periodic motions are ``mixed'' 
by noise resulting in 
new, global isophases. To this end we analyze the following model of two 
coexisting stable limit cycles, driven by white noise:
\begin{equation} \begin{aligned}
	\dot{r} &= r(1-r)(3-r)(c-r)+\sigma\xi(t)~,\\
	\dot{\theta} &= \omega+\delta(r-2)-(1-r)(3-r)~.
\end{aligned} \label{eq:bstc} \end{equation}
Without noise, the system shows two limit cycles $r_I=1,r_{II}=3$ (which 
have the same frequency
if $\delta=0$), separated by an unstable cycle at $r=c$. Each of the stable 
cycles has its own isophases, which meet singularly (as infinitely rotating
 spirals) at the basin boundary $r=c$. With noise, trajectory switches 
 between the basins, so that combined mixed-mode oscillations involving 
 both cycles occur. By applying our method, we find the isophases of these 
 oscillations in the whole range of the amplitudes, as shown in 
 Fig.~\ref{fig:doubleLS-cmp-isoph}. While for small noise amplitude
a residue of the singularity at the basin boundary is clearly seen, for a 
strong noise the isophases are rather smooth curves.

\begin{figure}[h]
\centering
\includegraphics[width=0.8\columnwidth]{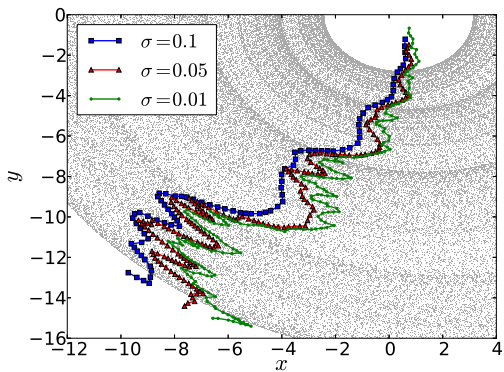}
	\caption{ (color online) Isophases of the noise-driven Roessler chaotic 
	system (\ref{eq:roes}), for  different noise intensities $\sigma$. 
	Smaller noise leads to less smooth curves. Grey dots show the 
	deterministic Roessler attractor.}
	\label{fig:roes}
\end{figure}

Another example where otherwise singular isophases are smeared by noise 
is that of chaotic oscillations. Many chaotic attractors allow a 
representation in terms of amplitudes and 
phases~\cite{Farmer-81,Pikovsky-85,Pikovsky2001}, but
because the phase generally performs a chaos-induced diffusion, isophases 
in the strict sense do not exist. Recently, description of chaotic 
oscillations in terms of approximate isophases has been 
suggested~\cite{Schwabedal2012}. With noise, the return times to a 
Poincar\'e surface of a strange attractor can be defined in the averaged 
sense only, and in this respect there is no difference between chaotic and 
regular deterministic oscillators. Thus, the procedure of finding isophases 
based on the constancy of the MFRTs can be applied to chaotic systems as 
well, as is illustrated in Fig.~\ref{fig:roes} for the Roessler model
\begin{equation}\begin{aligned}
	\dot{x} &=  -y-z+\sigma \xi_1(t)~, \\
	\dot{y} &=  x + 0.16y+\sigma \xi_2(t)~, \\
	\dot{z} &= 0.2+z(x-10)~,
\end{aligned}\label{eq:roes}\end{equation}
with uncorrelated Gaussian white noises in $x$ and $y$ components.

\begin{figure}[h]
	\centering
		\includegraphics[width=0.8\columnwidth]{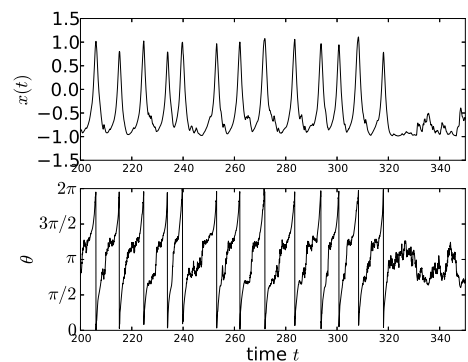}
	\caption{ At $\omega=1.99$, $\kappa=1$, and $\sigma=0.6$, system
	(\ref{eq:Lsex}) shows seemingly self-sustained oscillations (with some rather noisy patches as well), where indeed
	oscillations are noise-induced.  Top panel: variable $x(t)=r\cos\theta$, bottom panel: $\theta(t)$. }
	\label{fig:plt-proto-radius}
\end{figure}

\begin{figure}[h]
\centering
\includegraphics[width=0.8\columnwidth]{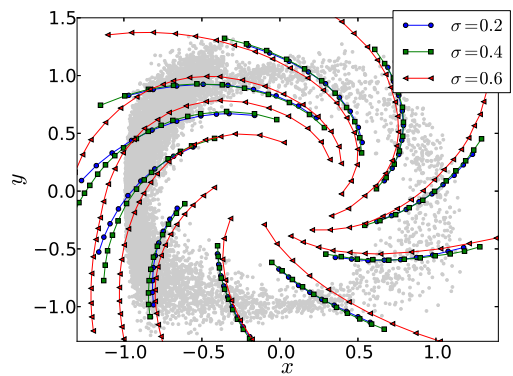}
\caption{ (color online). Average isophases of Eq.~(\ref{eq:Lsex}) at
	$\omega=1.99$ and $\kappa=1$ vary with noise intensity $\sigma$ as indicated.
	Larger noise intensity makes the system less isochronous, letting average
	isophases show stronger curvature. (Background trajectory of noise-induced oscillations (grey line) corresponds to the strong noise case $\sigma=0.6$.) }
\label{fig:plt-iso-together}
\end{figure}

Our final example are noise-induced oscillations in an excitable system, 
which without noise has just a stable steady state, so deterministic 
isophases do not exist in any sense. With noise, such a system demonstrates 
oscillations which may be quite regular in the case of coherence 
resonance~\cite{Pikovsky1997}.
To build the model, we modify the noisy Stuart-Landau oscillator, with $y$-
polarized noise, to perform
noise-induced oscillations:
\begin{equation}\begin{aligned}
	\dot{r} &= r(1-r^2)+\sigma r\cos\theta~\xi(t)~,\\
	\dot{\theta} &= \omega+r\cos\theta-\kappa r^2+\sigma\sin\theta~\xi(t)~.
\end{aligned}\label{eq:Lsex}\end{equation}
For $0<\omega-\kappa<1$ there is a stable state at $r_0=1,\theta_0=\pi-\arccos(\omega-\kappa)$ and an unstable state at $r_1=1,\theta_1=\pi+\arccos(\omega-\kappa)$ (at $\omega-\kappa=1$ 
they give rise 
to a periodic oscillation via a SNIPER bifurcation). 
Noise ($\sigma\neq0$) excites the state $r_0,\theta_0$ beyond  $r_1,\theta_1$ and produces noise-induced oscillations 
(Fig.~\ref{fig:plt-proto-radius}). For strong excitability and small noise, the phase is well-
defined, and the isophases can be introduced as curves with constant MFRTs. 
We show ten isophases in Fig.~\ref{fig:plt-iso-together} (examples 
Fig.~\ref{fig:colored-plt-poincare} and Fig.~\ref{fig:doubleLS-cmp-isoph} 
have been rotationally symmetric, so one drawing of one isophase was sufficient, 
here the rotational symmetry is broken).
The effect of the noise intensity on the isophases is maximal at 
$\theta\approx\pi$, i.e. in the region of excitability where the 
oscillations spend most of the time; 
in the ``deterministic'' region $|\theta|<\pi/2$ the isochrones are less sensitive to noise.

A practical definition of the isophases for which the MFRT is constant, is 
straightforward for numerical models of irregular oscillators as 
illustrated above, but it can be used for experimentally observed signals 
as well. For this purpose one needs a two-dimensional embedding of observed 
oscillations, which can be, e.g., achieved by using the Hilbert transform 
of the signal as the second variable. 
In Fig.~\ref{fig:resp} we present such a representation of measurements of 
human respiration, taken from the Physionet database~\footnote{The data (75 
oscillations, 66700 data points sampled at 0.004s, mean period 3.55s) is 
taken from the ``Fantasia database'' publicly available through
\textit{www.physionet.org};  the subject used is f1y01 (young adult, 
breathing
calmly/regular while watching the Walt Disney movie ``Fantasia'')}. One can 
see that the oscillations have a large amplitude variability, and defining 
the phase has a large degree of ambiguity -- contrary to the situations 
with a nearly constant amplitude, where a similar embedding on the 
signal vs its Hilbert transform plane results in a very narrow band of trajectories. 
 The initial phase-like variables and the isophases resulting from the 
 iterative procedure as described above are presented in 
 Fig.~\ref{fig:resp}. Application of the calculated isophases to 
 determining the
phase dynamics of the observed signals gives the mostly uniformly rotating 
phase, maximally uncorrelated from the amplitude variations.

\begin{figure}[h]
\centering
\includegraphics[width=\columnwidth]{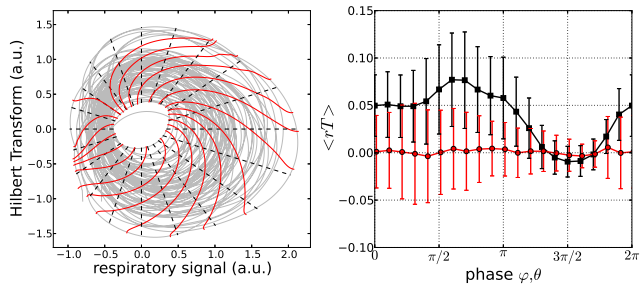}
\caption{ (color online). Left panel: Arbitrary phaselike variables (black) 
and isophases (red) of the experimental data of a respiration signal taken 
from the public database ``Physionet''. Right panel: for isophases the 
return times are independent of the radial variable (cross-correlation 
vanishes).}
\label{fig:resp}
\end{figure}

Summarizing, we have introduced for irregular oscillations a concept of 
average
isophases, based on the constancy of the mean first return times. By 
applying a simple procedure, we determined these isophases in a unified way
for different classes of noisy oscillators: (i) noise-perturbed periodic 
oscillators, which posses isophases also in the noise-free case; (ii) 
multistable oscillators which in the noise-free case posses different 
singular
isophases, but the latter become well-defined when different modes merge 
due to noise; (iii) chaotic attractors where in the purely  deterministic 
case the isochrones are singular objects which become smooth 
and well-defined 
due to noise; (iv) excitable systems which do not oscillate without noise 
and therefore have no isophases,
 but the latter appear for the noise-induced dynamics. Furthermore, we have 
 demonstrated applicability of the method to irregular experimental data. 
 The definition of isophases in noisy systems has two potential application 
 fields: in the data analysis, where it allows one to perform a consistent 
 phase reduction for signals with large amplitude variations, and in the 
 synchronization theory, serving at a determining of phase responses to 
 external kicks.

\begin{acknowledgments}

J.~S.~was partly supported by the DFG (Collaborative Research Project 555
``Complex Nonlinear Processes'').

\end{acknowledgments}

%

\end{document}